\documentclass{vldb}
\usepackage{graphicx}
\usepackage{balance}  

\usepackage{hyphenat}

\usepackage{amssymb,amsfonts}
\usepackage{graphicx}
\usepackage{textcomp}
\usepackage{xcolor}
\usepackage[hyphens]{url}
\usepackage{hyperref}
\hypersetup{breaklinks=true}
\usepackage{amsfonts}
\usepackage{amssymb}
\usepackage{multirow}
\usepackage{algorithm}
\usepackage{algpseudocode}
\usepackage{graphicx}
\usepackage{url}
\usepackage{subcaption}

\graphicspath{{figures/}}

\newtheorem{theorem}{Theorem}[section]
\newtheorem{lemma}{Lemma}[section]
\newtheorem{definition}{Definition}[section]

\usepackage{color,soul}

\usepackage{tikz}

\usepackage{environ}
\usepackage{setspace} 

\newenvironment{alignSmall}{\nobreak\small\noindent\align}{\endalign}
\newenvironment{alignFootnotesize}{\nobreak\footnotesize\noindent\align}{\endalign}
\newenvironment{alignScriptsize}{\nobreak\scriptsize\noindent\align}{\endalign}

\NewEnviron{myAlignS}{%
	\begin{singlespace}
		\vspace{-3pt}
		\begin{alignSmall}
			\BODY
		\end{alignSmall}
		\vspace{-5pt}
	\end{singlespace}
}

\NewEnviron{myAlignSS}{%
	\begin{singlespace}
		\vspace{-3pt}
		\begin{alignFootnotesize}
			\BODY
		\end{alignFootnotesize}
		\vspace{-5pt}
	\end{singlespace}
}

\NewEnviron{myAlignSSS}{%
	\begin{singlespace}
		\vspace{-5pt}
		\begin{alignScriptsize}
			\BODY
		\end{alignScriptsize}
		\vspace{-5pt}
	\end{singlespace}
}

\vldbTitle{PANDA: Policy-aware Location Privacy for Epidemic Surveillance}
\vldbAuthors{Yang Cao, Shun Takagi, Yonghui Xiao, Li Xiong,  Masatoshi Yoshikawa}
\vldbDOI{https://doi.org/10.14778/3352063.3352086}
\vldbVolume{12}
\vldbNumber{12}
\vldbYear{2019}

\newcommand\given[1][]{\:#1\vert\:}

\begin{document}


\title{PANDA: Policy-aware Location Privacy for Epidemic Surveillance}



%
%
%
%

\numberofauthors{5} 
\author{
	\alignauthor
	Yang Cao\\
	\affaddr{Kyoto University, Japan}\\
	\email{yang@i.kyoto-u.ac.jp}
	\alignauthor
	Shun Takagi\\
	\affaddr{Kyoto University, Japan}\\
	\email{takagi.shun.45a@st.kyoto-u.ac.jp}
		\alignauthor
	Yonghui Xiao\\
	\affaddr{Google Inc., USA}\\
	\email{yohu@google.com}
	\and
		\alignauthor
	Li Xiong\\
	\affaddr{Emory University, USA}\\
	\email{lxiong@emory.edu}
	\alignauthor
Masatoshi Yoshikawa\\
\affaddr{Kyoto University, Japan}\\
\email{yoshikawa@i.kyoto-u.ac.jp}
}

\maketitle

\begin{abstract}
In this demonstration, we present a  privacy-preserving epidemic surveillance system.
Recently, many countries that suffer from coronavirus crises attempt to access citizen's location data to eliminate the outbreak.
However, it raises privacy concerns and may open the doors to more invasive forms of surveillance in the name of public health.
It also brings a challenge for privacy protection techniques: how can we leverage people's mobile data to help combat the pandemic without scarifying our location privacy.
We demonstrate that we can have the best of the two worlds by implementing policy-based location privacy for epidemic surveillance.
Specifically, we formalize the privacy policy using graphs in light of differential privacy, called policy graph.
Our system has three primary functions for epidemic surveillance: location monitoring, epidemic analysis, and contact tracing.
We provide an interactive tool allowing the attendees to explore and examine the usability of our system:
(1) the utility of location monitor and disease transmission model estimation,
(2) the procedure of contact tracing in our systems,
and (3) the privacy-utility trade-offs w.r.t. different policy graphs.
The attendees can find that it is possible to have the full functionality of epidemic surveillance while preserving location privacy.
\end{abstract}

\section{Introduction}
\label{sec-intro}

We are fighting with the pandemic of COVID-19 disease.
To prevent the spread of such a highly contagious virus, the crucial information that we need is people's location history for epidemic surveillance. 
Recently, many countries that suffer from coronavirus crises attempt to access citizen's location data to eliminate the outbreak.
The US pumped 500 million dollars of emergency funding into the CDC for building a surveillance and data collection system \cite{web1} and discussed with Facebook and Google for sharing people's location data to combat the coronavirus.
In South Korea, the government created a public map of coronavirus patients using location data from telecom and credit card companies \cite{web2}.
Italy's telecom companies are sharing location data with health authorities to check whether people are remaining at home \cite{web3}.
China's giant tech companies provide a ``health code'' service to certificate a user's health status based on her health status and travel history, which are collected by the cellphone Apps \cite{web4}.
Although these special measures of personal data collection for public health emergency may be temporary and under stringent government regulation,  it raises concerns over privacy, and people are worried that it may open the doors to surveillance activities in the name of public health. 
It also brings a challenge for location privacy protection techniques: how can we utilize people's mobile data to help combat the pandemic without sacrificing our location privacy.

Location privacy has been extensively studied in the literature  \cite{primault_long_2018}.
However, the state-of-the-art  location privacy models are not flexible enough to balance the individual privacy and public interest in an emergency as we are witnessing in the COVID-19 crisis.
The early studies on location privacy were extending  $ k $-anonymity \cite{sweeney_k-anonymity:_2002}  and were flexible enough to be adapted to different scenarios such as personalized location anonymity \cite{gedik_protecting_2008}.
But the recent studies revealed that $ k $-anonymity might not be rigorous enough since they suffer many realistic attacks \cite{li_t-closeness:_2007, machanavajjhala_l-diversity:_2006} when the adversary has background knowledge about the original dataset.
The recent state-of-the-art location privacy models\cite{geoi_ccs13, xiao_ccs15, theodorakopoulos_prolonging_2014,  takagi_GGI_2019, xiao_loclok:_2017, cao_priste:_2019, cao_priste:_2019-1, cao_protecting_2019} were extended from differential privacy (DP) \cite{Dwork06differentialprivacy}  to private location release since DP is considered a rigorous privacy notion.
Although these DP-based location privacy models are rigorously defined, yet they are not flexible and customizable for different scenarios with various requirements on {privacy-utility trade-off}. 
Taking an example of Geo-Indistinguishability\cite{geoi_ccs13}, which is the first and influential DP-based location privacy metrics, the strength of protection is solely controlled by a single parameter $ \epsilon $ to achieve indistinguishability among all possible locations.
It is hard to make a good privacy-utility trade-off using this single $ \epsilon $ in a complicated setting. 

We should have a flexible and rigorous location privacy model that enables customizable location privacy policy, which defines which locations are sensitive, which are not.
The policy should be adjustable for different people, at different time, and with different purposes.
For instance, under the emergency of COVID-19,  a location privacy policy for contact tracing could be  ``\textit{allowing to disclose a user's true locations of the past two weeks if she is a diagnosed coronavirus patient; otherwise, ensuring indistinguishability of the user's location}'';
if the patient's location trace and the time period are confirmed, we can dynamically update the location privacy policy for each person to find all contacts of the confirmed patient.
A policy for all other people could be ``\textit{allowing to disclose a user's true locations if she has been stay in the same location at the same period; otherwise, ensuring indistinguishability of the user's location}''
In this way, we can guarantee both full usability of contact tracing and reasonable privacy.

\begin{figure}[t]
	\centering
	\includegraphics[width=6cm]{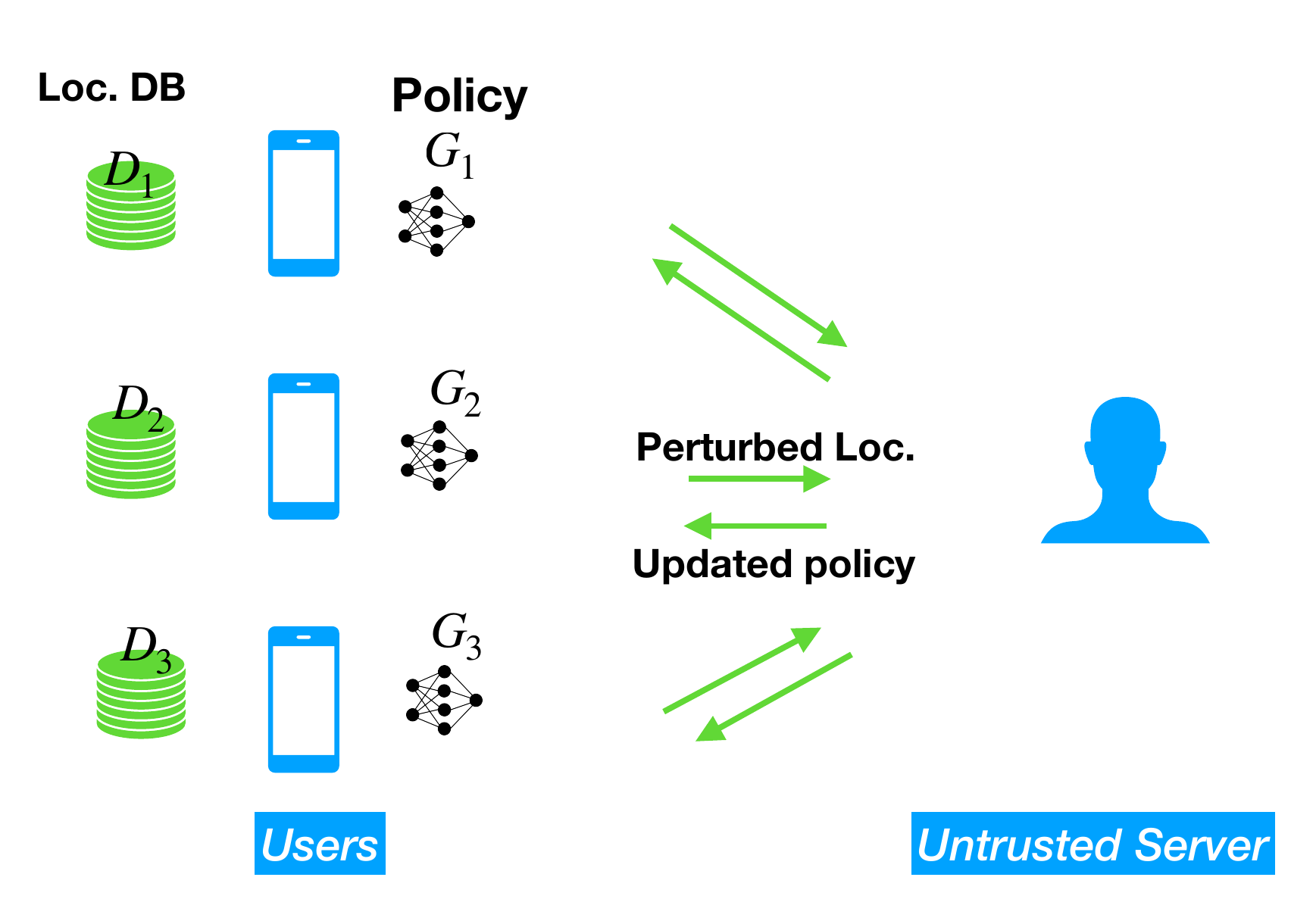}
	\vspace{-10pt}
	\caption{Private location sharing with Customizable Policy.}
	\label{fig1}
	\vspace{-15pt}
\end{figure}

In this demonstration, we present PANDA, i.e., \underline{P}olicy-aware priv\underline{A}cy preservi\underline{N}g epiDemic surveill\underline{A}nce, which implements our recently proposed Policy Graph-based Location Privacy (PGLP) \cite{cao20} and mechanisms for epidemic surveillance.
Our system is featured by the customizable location privacy policy graph, which provides a new dimension to tune utility-privacy trade-off.

In our recent study \cite{cao20}, we proposed a formal representation of location privacy policy using a graph, which is inspired by a statistical privacy notion of Blowfish privacy \cite{Blowfish-SIGMOD14}.
In our setting of private location release, a privacy policy graph  (such as the ones shown in Fig.\ref{figure-two-graphs}) includes all possible locations that need to be protected as its nodes, and the edges indicate indistinguishability between two possible locations.
A user could arbitrarily customize the location policy graph according to her privacexy and utility requirement and enjoy plausible deniability regarding her whereabouts.
The definition of PGLP can be seen as a generalization of two influential  DP-based location privacy models: Geo-Indistinguishability \cite{geoi_ccs13} and Location Set Privacy \cite{xiao_ccs15}.
Under appropriate configuration of policy graphs, an algorithm satisfying PGLP w.r.t. the policy graphs could also satisfy  Geo-Indistinguishability or Location Set Privacy.
In \cite{cao20}, we also designed mechanisms for PGLP by adapting the Laplace mechanism and Planar Isotropic Mechanism (PIM) (i.e., an optimal mechanism for Location Set Privacy  \cite{xiao_ccs15}) w.r.t. a given location policy graph.

However, it is not trivial to directly apply PGLP for a location-based application such as epidemic surveillance due to the following reasons.
First, it is not clear how to design a proper policy graph with reasonable privacy and functional utility.
Second, when there are multiple choices for location privacy policies, we lack a tool to explore and compare the utility gain w.r.t. different location privacy policies.
Third, it is difficult for users to understand the privacy implications (i.e., the privacy risks) of a given location privacy policy.

\subsection{Contributions}
To address the above issues and motivated by the significant impact of the pandemic of COVID-19 in the world, we demonstrate a policy-based location privacy-preserving epidemic surveillance system.
Our contributions are summarized below.

First, we design an epidemic surveillance system with three primary functions:\textit{ location monitoring, epidemic analysis}, and \textit{contact tracing}.
The scenario is shown in Fig.\ref{fig1}, where users locally maintain location databases (e.g.,  all locations in the past two weeks) and share perturbed locations satisfying PGLP w.r.t. a specific policy graph with a semi-honest server.
The policy graph essentially acts as an information filter to control what could be shared and what should not be shared.

Second, we demonstrate three policy graphs with the distinct granularity that are appropriate for different functions in the epidemic surveillance.
Specifically, we visualize the utility gain or loss between different policy graphs.
It turns out that no policy could be the best for all.
The attendees of the conference can find that it is possible to have the full functionality of epidemic surveillance while preserving location privacy.

Third, we visualize the trade-off between privacy and utility.
Although we can specify a policy graph that enables the full usability of the system, yet it is not clear what is the privacy implication given a policy graph. 
The policy graph itself could be semantically meaningful, but we lack a quantitative measurement.
We provide empirical privacy metrics  as the adversary's successful inference \cite{Quantifying-location-privacy-SP2011} with an interactive tool 
The attendees can randomly generate a policy graph to explore its effect on the privacy-utility trade-off.
The code is available in github\footnote{\url{https://github.com/tkgsn/pglp}.}.
A  prototype of a mobile phone App will be available soon.

\section{Background}

\subsection{Location Policy Graph}
\label{subsec-lpg}

Inspired by Blowfish privacy\cite{Blowfish-SIGMOD14}, we use an undirected graph to define which location should be protected and which could not, i.e., location privacy policies.
The nodes are secrets and the edges are the required indistinguishability, which indicate an attacker should not be able to distinguish the input secrets by observing the perturbed output.  
In our setting, we treat possible locations as nodes, and the indistinguishability between the locations as edges.

\begin{definition}[Location Policy Graph]
	A location policy graph is an undirected graph $\mathcal{G}=(\mathcal{S},\mathcal{E})$ where
	$\mathcal{S}$ denotes all the locations (nodes) and $\mathcal{E}$ represents indistinguishability (edges) between these locations.
\end{definition}

\begin{definition}[Distance in Policy Graph]
We define the distance between two nodes $\textbf{s}_i$ and $\textbf{s}_j$ in a policy graph as the length of the shortest path between them, denoted by $d_\mathcal{G}(\textbf{s}_i, \textbf{s}_j)$.
\end{definition}

In DP, the two possible database instances with or without a user's data are called  \textit{neighboring databases}, which can be interpreted as two nodes with an edge in a policy graph.
We  generalize it to \textit{k-neighbors} defined below.

\begin{definition}[k-Neighbors]
\label{def-neighbors}
	The k-neighbors of location $\textbf{s}$, denoted by $\mathcal{N}^k(\textbf{s})$, is the set of nodes that reach $\textbf{s}$ within k hops, i.e.,	$\mathcal{N}^k(\textbf{s})= \{\textbf{s}' \given d_\mathcal{G}(\textbf{s},\textbf{s}') \leq k, \textbf{s}'\in \mathcal{S}\}$.
	We define  $\infty$-neighbors as the nodes having a path with  $\textbf{s}$, denoted by $\mathcal{N}^\infty (\textbf{s})$.
\end{definition}

In our system, we assume that the location policy graph is determined by the server for the purposed of utility maximization.
The user has the right to reject a privacy policy so that no location will be released.
By making the policy graph public, the system has a high level of transparency.

\subsection{Privacy Metrics}
\label{subsec-metrics}


We now formalize PGLP (i.e., Policy-based Location Privacy), which guarantees indistinguishability for every pair of neighbors (i.e., for each edge) in a  location policy graph.

\begin{definition}[$\{\epsilon,\mathcal{G}\}$-Location Privacy]
	\label{def-DPMC}
	A randomized algorithm $\mathcal{A}$  satisfies $\{\epsilon,\mathcal{G}\}$-location privacy iff for all $  \textbf{z} \subseteq Range(\mathcal{A})$ and for all pairs of $1$-neighbors $   \textbf{s} $ and $  \textbf{s}'$ in $\mathcal{G}$, we have   {\small $
	\frac{\Pr(\mathcal{A}(\textbf{s})=\textbf{z} )}{\Pr(\mathcal{A}(\textbf{s}')=\textbf{z} )}\leq e^{\epsilon}
$}.
\end{definition}

In PGLP,  privacy is rigorously guaranteed through ensuring indistinguishability between any two neighboring locations specified by a customizable location policy graph.
 The user enjoys plausible deniability about her whereabout.

\begin{lemma}
\label{lm-graph-dist}
An algorithm $\mathcal{A}$ satisfies $\{\epsilon,\mathcal{G}\}$-location privacy, iff any two $\infty$-neighbors $\textbf{s}_i, \textbf{s}_j\in {\mathcal{G}}$ are $ \epsilon \cdot d_\mathcal{G}(\textbf{s}_i, \textbf{s}_j)$-indistinguishable.
\end{lemma}

Lemma \ref{lm-graph-dist} indicates that,  if there is a path between two nodes (locations) $\textbf{s}_i, \textbf{s}_j$ in the policy graph, the corresponding indistinguishability is required at a certain degree; if two nodes are not connected (i.e., $d_\mathcal{G}(\textbf{s}_i, \textbf{s}_j) = \infty$), the indistinguishability  is not required by the policy.
As an extreme case, if a node is not connected with any other nodes, it allows to release it without any perturbation.

\begin{figure}[t]
	\centering
	\includegraphics[width=8cm]{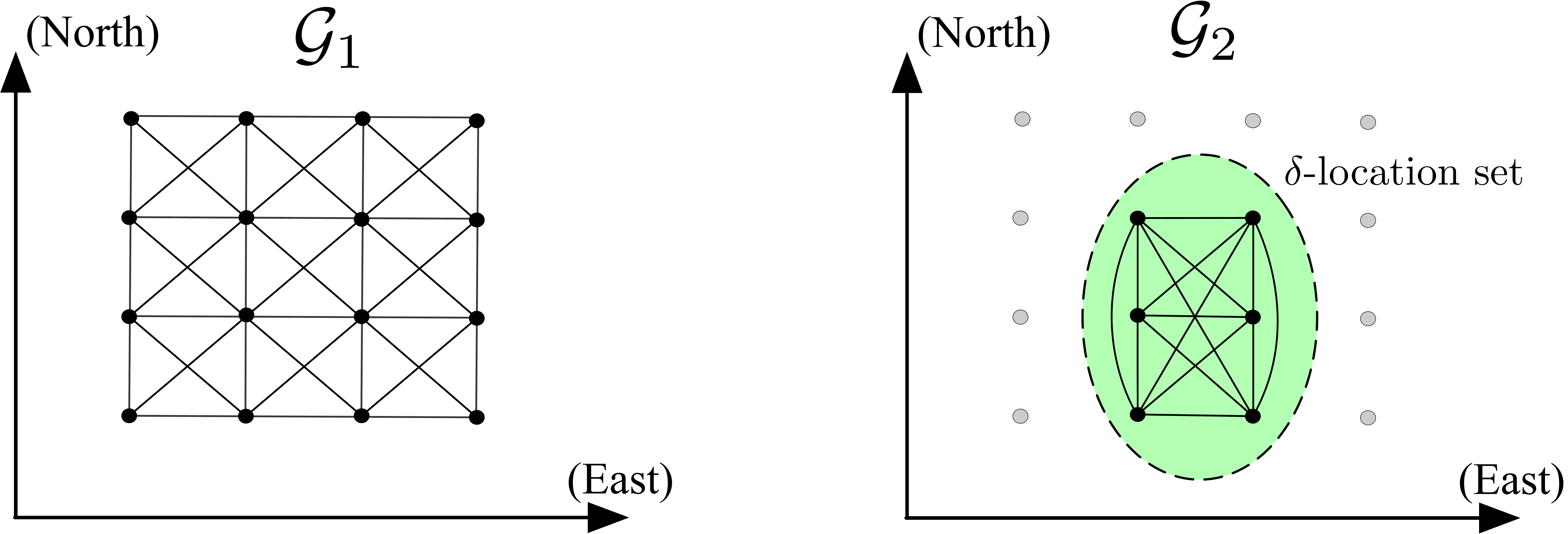}
	\vspace{-5pt}
	\caption{Two examples of location policy graphs.}
	\label{figure-two-graphs}
		\vspace{-15pt}
\end{figure}

\subsubsection{Comparison with Other Location Privacy}
We analyze the relation between PGLP and two influential DP-based location privacy models, i.e., {Geo-Indistinguishability} \cite{geoi_ccs13} and $\delta$-Location Set Privacy \cite{xiao_ccs15}. 
We show that PGLP implies each of them under proper configurations of location policy graphs.

{\textit{Geo-Indistinguishability} \cite{geoi_ccs13}} guarantees a level of indistinguishability between two locations $\textbf{s}_i$ and $\textbf{s}_j$ that is scaled with their Euclidean distance, i.e.,  $\epsilon \cdot d_E(\textbf{s}_i, \textbf{s}_j) $-indistinguisha-bility, where $d_E(\cdot, \cdot )$ denotes  Euclidean distance. 
Let $\mathcal{G}_1$ be a location policy graph that every location has edges with its closest eight locations on the map as shown in Fig.\ref{figure-two-graphs} (left). 
We can derive the following theorem by the fact of $d_\mathcal{G}(\textbf{s}_i, \textbf{s}_j) \leq d_E(\textbf{s}_i, \textbf{s}_j)$ for any $\textbf{s}_i, \textbf{s}_j \in \mathcal{G}_1$ and Lemma \ref{lm-graph-dist}.

\begin{theorem}
An algorithm satisfying $\{\epsilon, \mathcal{G}_1\}$-location privacy also achieves $\epsilon$-Geo-Indistinguishability.
\end{theorem}

\textit{$\delta$-Location Set Privacy} \cite{xiao_ccs15} extends differential privacy on a subset of possible locations, which is assumed as adversarial knowledge.
$\delta$-Location Set Privacy ensures indistinguishability among any two locations in the $\delta$-location set.
Let $\mathcal{G}_2$ be a location policy  that is a complete graph among locations in the $\delta$-location set as shown in Fig.\ref{figure-two-graphs} (right).

\begin{theorem}
An algorithm satisfying $\{\epsilon, \mathcal{G}_2\}$-location privacy also achieves $\delta$-Location Set privacy. 
\end{theorem}

The proofs and the mechanisms for PGLP are presented in a full version of this paper \cite{cao20} for  interested readers.

\section{System Overview}

\begin{figure}[t]
	\centering
	\begin{subfigure}{0.65\textwidth}
		\includegraphics[width=8cm]{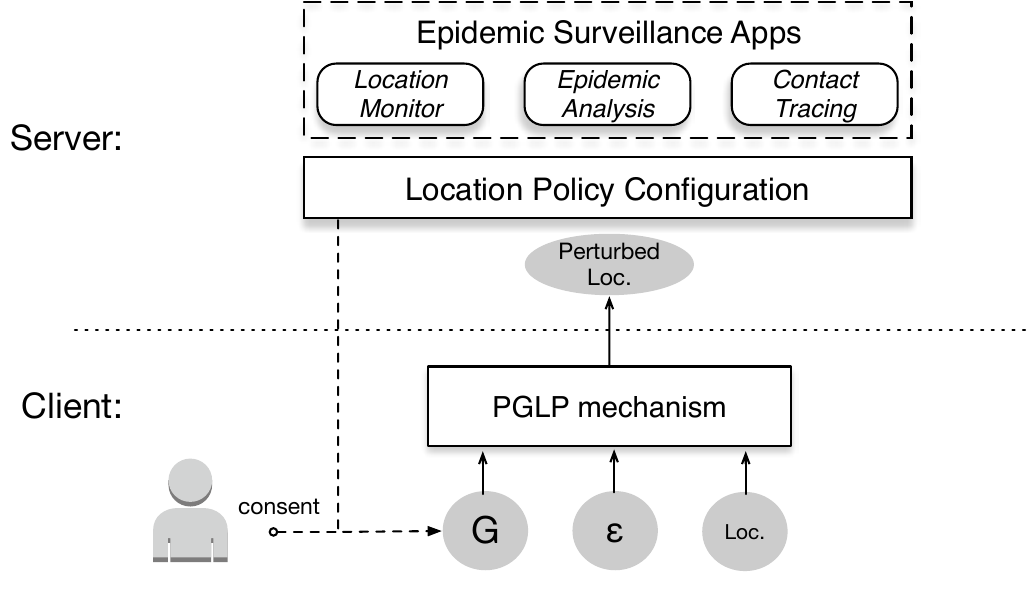}
	\end{subfigure}
\vspace{-15pt}
	\caption{System Overview.}
	\label{fig:sys}
	\vspace{-15pt}
\end{figure}

\subsection{Epidemic Surveillance}

\begin{figure}[b]
	\centering
	\begin{subfigure}{0.42\textwidth}
		\includegraphics[width=8cm]{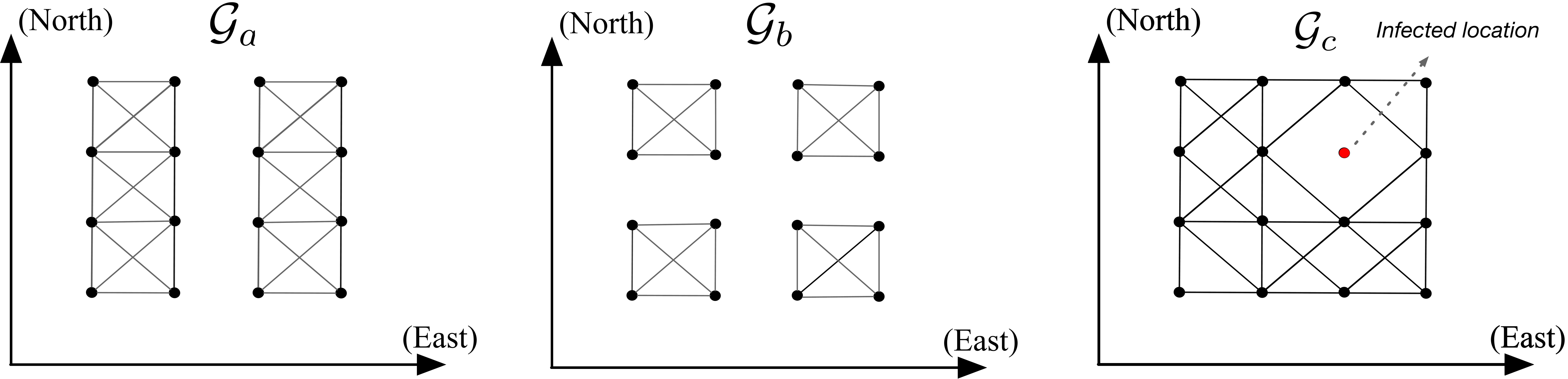}
	\end{subfigure}
	\vspace{-10pt}
	\caption{Location policy graphs for epidemic surveillance.}
	\label{fig:3g}
	\vspace{-10pt}
\end{figure}

Our system provides consist of three main modules: PGLP mechanisms,  Location Policy  Configuration, and Epidemic Surveillance Apps as shown in Fig.\ref{fig:sys}.
PGLP mechanisms are proposed in \cite{cao20} for achieving rigorous and customization location privacy.
It takes inputs of $ \epsilon $, location policy graph $ G $ and the user's true location, and outputs a perturbed location to the server.
The policy $ G $  recommended by Location Policy  Configuration and approved by the user.
Location Policy  Configuration defines different location policies according to the application of epidemic surveillance.
Three primary functions (Apps) for epidemic surveillance are location monitoring, epidemic analysis and contact tracing.
\textit{Location monitoring}  focuses on understanding people's movement between different cities or provinces in a coarse-grained level, which provides essential insights when combining with the incidence rate in each city along with the people's movement.
It could also provide a ``health code'' service, i.e.,  allowing certification of the user’s health status, in a privacy-preserving way.
A location policy for location monitoring can be ``\textit{ensuring indistinguishability inside each coarse-grained area and allowing the locations are distinguishable in different coarse-grained areas}'' such as $ \mathcal{G}_a $ shown in Fig.\ref{fig:3g} since such a monitor only requires the people moving between different cities.
\textit{Epidemic analysis}  aims at building a predictive disease transmission model such as the SEIR model \cite{li1995global}.
The fine-grained data would be beneficial for the estimation of the parameters such as R0 (i.e., basic reproduction number). 
A location policy for epidemic analysis is similar to the previous one, but more fine-grained, such as $ \mathcal{G}_b $  in Fig.\ref{fig:3g}.
\textit{Contact tracing}  attempts to find all contacts of a diagnosed case so that to stop the spread of disease by finding and isolating patients.
A policy for contact tracing can be ``ensuring indistinguishability only if the user is not in an infected area, but allowing disclose true location if the user accesses an infected location'', which can be formally represented by a graph  $ \mathcal{G}_c $  in Fig.\ref{fig:3g}. 
We introduce more details about contact tracing below.

\begin{figure}[t]
	\centering
	\begin{subfigure}{0.8\textwidth}
		\includegraphics[width=7cm]{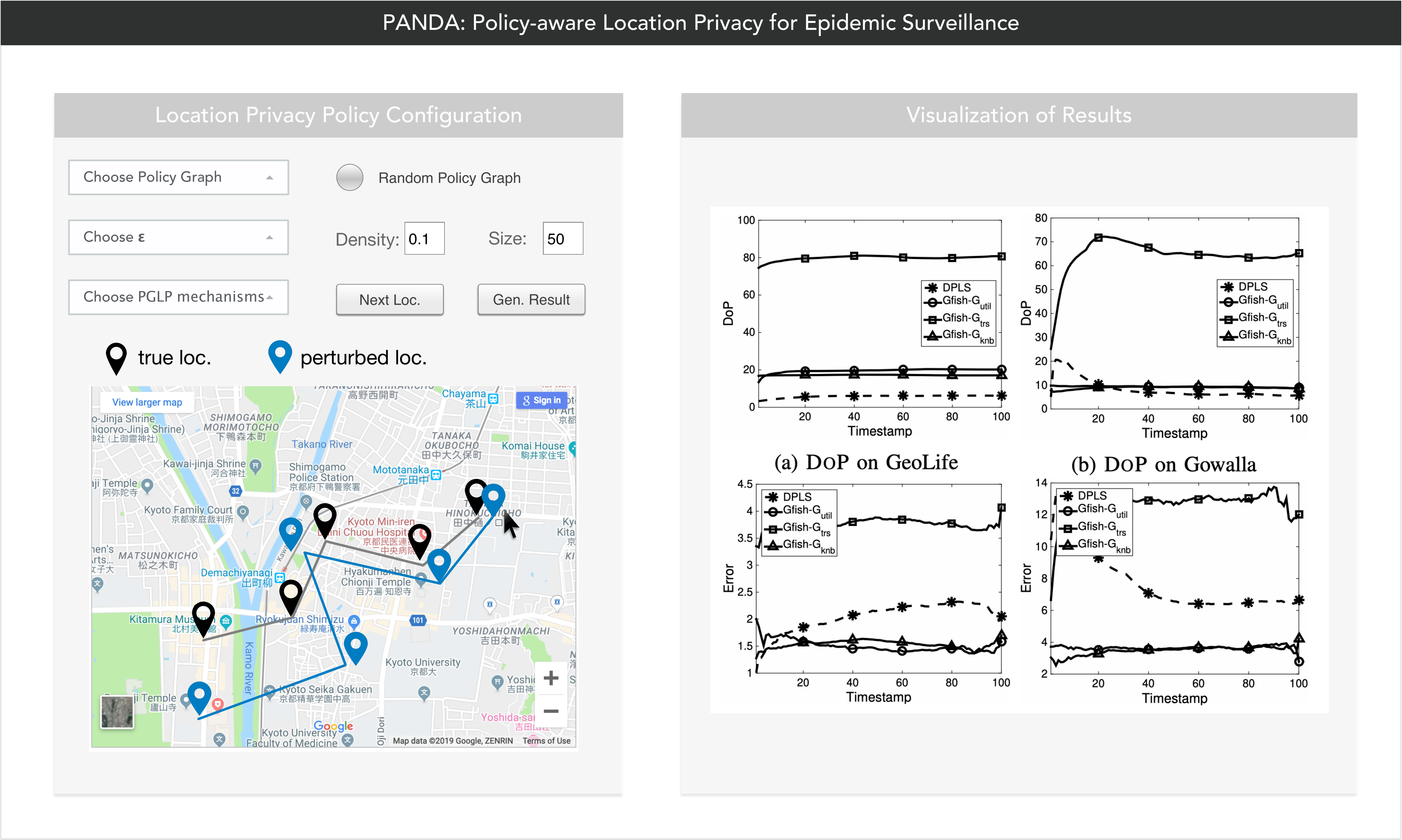}
	\end{subfigure}
	\vspace{-5pt}
	\caption{PANDA Demonstration.}
	\label{fig:ex1}
	\vspace{-15pt}
\end{figure}

\subsection{Demonstration Scenario}

We demonstrate the system using  Geolife  \cite{GeoLife-Zheng-10}  and Gowalla \cite{KDD-Gowalla-2011}  datasets.
Interested readers can find a more detailed configuration in \cite{cao20}.
We provide an interactive tool that allows the attendees to explore and examine the usability of our system:
(1) the utility of location monitor and coronavirus transmission model estimation,
(2) the procedure of contact tracing in our systems,
and (3) the privacy-utility trade-offs, as shown in Fig.\ref{fig:ex1} w.r.t. different policy graphs.
First, we evaluate the utility of location monitoring as the Euclidean distance between perturbed locations and real locations. 
We test the accuracy of transmission model estimation using the difference between (i.e., basic reproduction number) R0 estimated over accurate locations and the perturbed locations, respectively.
Second, we demonstrate the procedure of contact tracing using our system and dynamic policy graphs (such as $ \mathcal{G}_c $ in Fig.\ref{fig:3g}).
The goal is identifying the people who have the risk of infection (the decision rule of suspected infection could be advised by CDC or WHO; here we assume a simple rule of two persons have been the same location at the same time at least twice).
At each time point, each user sends the perturbed location w.r.t. her policy graph and stores the past two weeks of location history in a local database.
When the server confirms a diagnosed patient's location history, the Policy Graph Configuration module will update the location privacy policy of the users who have the risk of infection during the past two weeks (according to our simple rule).
Then, the corresponding user will be asked to re-send his past location using the updated privacy policy (the places where the diagnosed patient has been are allowed to be disclosed).
In this way, the user can get alerted and tested in case of infection.
Third, similar to the previous utility evaluation, we will also allow the attendees to evaluate the empirical privacy that is measured by adversary error \cite{Quantifying-location-privacy-SP2011}.
One can choose predefined policy graphs, as shown in Fig.\ref{fig:3g}, or randomly generate policy graphs to explore its effect on the privacy-utility trade-off.

\section{Acknowledgement}
This work is partially supported by JSPS KAKENHI Grant No. 17H06099, 18H04093, 19K20269, and Microsoft Research Asia (CORE16).




 \bibliographystyle{abbrv}

{\scriptsize{
\bibliography{ref}
}
\balance

\end{document}